# Dealing with missing angular sections in nanoCT reconstructions of low contrast polymeric samples employing a mechanical *in situ* loading stage

Short head: Dealing with missing angular sections in *in situ* nanoCT

Rafaela Debastiani[1,2*], Chantal Miriam Kurpiers[3], Enrico Domenico Lemma[7,8], Ben Breitung[1,2], Martin Bastmeyer[7,9], Ruth Schwaiger[4], Peter Gumbsch[1,5,6]

[1] Institute of Nanotechnology, Karlsruhe Institute of Technology, 76131 Karlsruhe, Germany

[2] Karlsruhe Nano Micro Facility (KNMFi), 76344, Eggenstein-Leopoldshafen, Germany

[3] Institute for Applied Materials - Mechanics of Materials and Interfaces, Karlsruhe Institute of Technology, 76344 Eggenstein-Leopoldshafen, Germany

[4] Institute of Energy and Climate Research - Microstructure and Properties of Materials (IEK-2), Forschungszentrum Jülich, 52425, Jülich, Germany

[5] Institute for Applied Materials - Reliability and Microstructure, Karlsruhe Institute of Technology, 76128, Karlsruhe, Germany

[6] Fraunhofer Institute for Mechanics of Materials IWM, 79108, Freiburg, Germany

[7] Zoological Institute, Karlsruhe Institute of Technology, 76131 Karlsruhe, Germany

[8] Department of Engineering, Università Campus Bio-Medico of Rome, 00128 Rome, Italy

[9] Institute of Biological and Chemical Systems - Biological Information Processing, Karlsruhe Institute of Technology, 76344 Eggenstein-Leopoldshafen, Germany

*Corresponding author:    R. Debastiani

rafaela.debastiani@kit.edu

+49 721 608 26358

Institute of Nanotechnology, Karlsruhe Institute of Technology,

P.O. Box 3640, 76021 Karlsruhe, Germany

**Abstract**

While *in situ* experiments are gaining importance for the (mechanical) assessment of metamaterials or materials with complex microstructures, imaging conditions in such experiments are often challenging. The lab-based computed tomography system Xradia 810 Ultra allows for the *in situ* (time lapsed) mechanical testing of samples. However, the *in situ* loading setup from this system limits the image acquisition angle to 140°. For low contrast polymeric materials, this limited acquisition angle leads to regions of low information gain, thus preventing an accurate reconstruction of the data using a filtered back projection algorithm. Here we demonstrate how the information gain can be improved by selecting an appropriate position of the sample. A low contrast polymeric tetrahedral microlattice sample and a specifically structured sample, both scanned over 140° and 180°, demonstrate that the missing structural details in the 140° reconstruction are limited to an angular wedge of about 20°. Depending on the sample geometry and structure, applying simple strategies for the *in situ* experiments allows accurate reconstruction of the data. For the tetrahedral microlattice, a simple rotation of the sample by 90° provides enough X-ray absorption for an accurate reconstruction of the geometry.

# 1. Introduction

*In situ* experiments are increasingly important for the understanding of materials behavior and properties. Combining experiments with X-ray computed tomography allows for the study of internal structural changes in a specimen as a function of time and applied stimulus. Among standard environmental setups integrated with X-ray imaging are heating devices [1,2] and mechanical loading devices [3–7]. X-ray computed tomography (CT) configurations vary in terms of sample size, imaging resolution and tomography acquisition speed. These CT parameters determine the kind of phenomena that can be observed in *in situ* experiments [8]. For laboratory-based X-ray computed tomography, the acquisition of a complete tomography may takes several hours, and thus *in situ* experiments are well suited to study processes that can be monitored in a time-lapsed manner, such as the formation of cracks upon stepwise loading [9]. For faster processes such as liquid metal foaming and sub-micron resolution, currently only synchrotron sources offer the required X-ray flux for the acquisition of hundreds of tomograms per second [10].

The laboratory-based transmission X-ray microscope Xradia 810 Ultra (nanoCT) offers the possibility of mechanical *in situ* testing under indentation, uniaxial compression or tension. Imaging is possible with a resolution down to 50 nm in absorption and Zernike phase contrast modes. While it is a great opportunity for imaging specimen under load with high resolution in a lab-based system, this particular setup also imposes a few limitations such as the maximum field of view of 65 μm, and a limited angle of 140° for the projections acquisition using the *in situ* load stage. With the load stage at a fixed position, the sample rotation is carried out inflexibly between -70° and +70°. The remaining +-20° are shadowed by the anvil.

Filtered back projection (FBP) is the most commonly used analytical algorithm to reconstruct computed tomography data [8] and the base for the Zeiss proprietary software Scout and Scan Reconstructor. The FBP reconstruction is based on the acquisition of 2D projections collected with equal angle increments over 180° or more. For each projection angle, the detector collects X-ray photographs of the sample. In the case of the nanoCT, which has a parallel beam geometry, each row of the detector can be reconstructed independently. The back projection algorithm then projects evenly the intensity of the X-ray photons collected by the detector along the angle in which it was recorded. With the collection of projections at different angles, the different projections intersect, thereby reconstructing the image of an object. Due to the uneven sampling at the center and at the edges of an object, the final back projection image is blurry. Applying a filter (e.g., ramp filter) to the projections suppresses the low frequency components in the Fourier space, compensating for the high frequency components missing due to the insufficient sampling, thus creating a sharper image. The reconstruction with FBP presents some limitations such as noise and image artefacts. Acquiring projections over a range smaller than 180°, as it happens when using the *in situ* load stage in the nanoCT, can result in areas of missing information in the reconstructed 3D image using FBP [8,11]. The missing information is particularly critical for samples with regions of low X-ray absorption.

In this article, we identify the inaccuracy in the reconstruction data of low absorption contrast samples scanned over a limited angle range using the nanoCT Xradia 810 Ultra equipped with the *in situ* load stage setup. Two polymeric specimens of different geometries were used to demonstrate the missing angular wedge in the reconstruction. A correct positioning of the sample can help to collect sufficient information to obtain an accurate reconstruction.

## 2. Materials and Methods

### 2.1. Sample preparation

Two samples denoted as samples A and B were printed using 3D direct laser writing (3D-DLW - Photonic Professional PPGT2, Nanoscribe GmbH) using IP-Dip resin (Nanoscribe GmbH). To avoid further manipulation of the samples, the polymeric samples were printed directly on the titanium pin of the *in situ* load stage.

For sample A, a tetrahedral microlattice geometry (Figure 1a) was chosen and printed with a laser power of 10.6 mW and a printing speed of 3000 µm/s as described in [12]. In addition, three bars were added to the geometry sticking out on the side of the sample (marked by arrows in Figure 1a) to serve as markers for the nanoCT scan. Sample B was a more uniform sample, created as a 50 μm diameter cylinder composed of vertical plates with a width 3 μm and different symbols at the end of the plates (Figure 1b). Here, the laser power was 25 mW and the scanning speed 5000 µm/s and further developed as described in [13].

After sample B was first scanned in the nanoCT, it was coated with a layer of $Al_2O_3$ to enhance the contrast and scanned again. The coating was deposited by Atomic Layer Deposition (ALD) using a Picosun R-200 Advanced system. A 100 nm thick alumina layer was prepared from trimethylaluminum (TMA) and H2O at 130 °C in 1750 cycles.

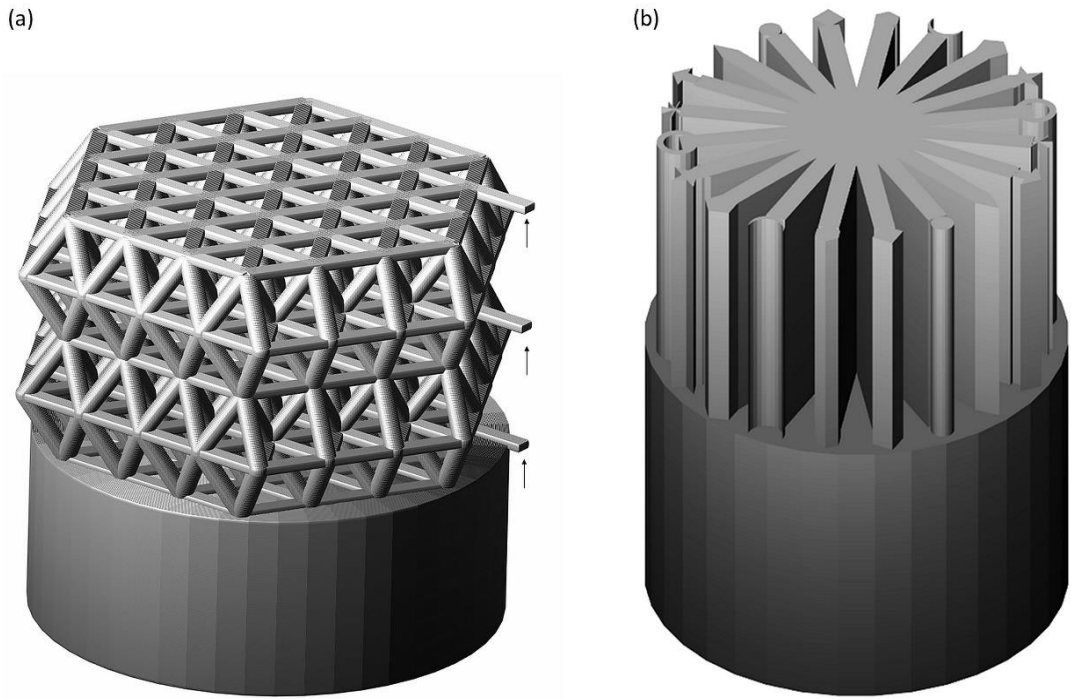

Figure 1: 3D models of the samples. (a) Sample A (diameter of 60 μm), a tetrahedral microlattice geometry with 3 bars sticking out on the side to serve as markers for the nanoCT scan (indicated by arrows). (b) Sample B (diameter of 50 μm), a cylindrical uniform sample with plates and different symbols at the end of the plates.

### 2.2. NanoCT

Samples A and B were scanned using the lab-based transmission X-ray microscope Zeiss Xradia 810 Ultra, referred to as nanoCT, for the acquisition of their 3D reconstruction. This system operates with a rotating Cr-anode (energy of 5.4 keV) and the samples in the current study were scanned in a field of view corresponding to 65 μm and pixel size of 128 nm. Absorption and Zernike phase contrast modes were employed to scan the samples using 901 and 1601 projections.

In order to understand the effects of angular range and the regions of low information gain when using the *in situ* loading setup provided by Zeiss (Figure 2a), experiments were performed with both the standard (Figure 2b) and *in situ* setup. The standard scan was carried out over 180° (maximum angle of projection acquisition in this system), from -90° to +90°, while the *in situ* scan was carried out from -70° to +70° (over 140°). In addition, the sample position for sample A was rotated to 45° and 90° (Figure 4a - pen mark on the side of the pin), with 0° corresponding to the typical sample positioning. Sample B was scanned at 0° and 90°. With the rotation, we aimed to verify the influence of the X-ray illumination of different parts of the sample on the data reconstruction when dealing with limited angle of scanning.

The projections were 3D reconstructed using the proprietary Zeiss Scout and Scan™ Control System Reconstructor (version 13.08) software, which is available with the equipment and is based on the filtered back projection algorithm. The reconstructed data was visualized using the ORS Dragonfly software [14].

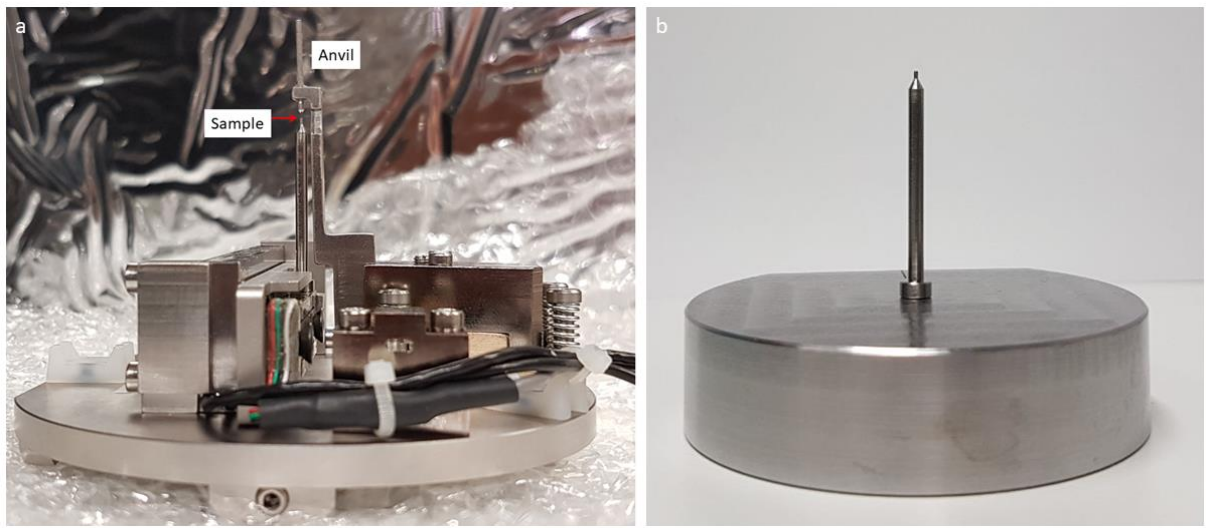

Figure 2: a) *in situ* loading stage from Xradia 810 Ultra in the compression and indentation setup. b) Standard stage from Xradia 810 Ultra.

## 3. Results

Using the mechanical *in situ* load stage for experiments in the nanoCT, the projection acquisition limited to 140° may cause the absence of sample information during the reconstruction. We observed such incorrect reconstruction in metamaterials. For example, the reconstruction of sample A scanned using the *in situ* setup showed the sample as defective, with the absence of the "horizontal beams" (Figure 3a). If sample A is scanned in the standard sample stage over 180°, the complete sample can be reconstructed (Figure 3b). This indicates that the missing beams were the result of the limited angle of acquisition and the weak X-ray absorption in such structures.

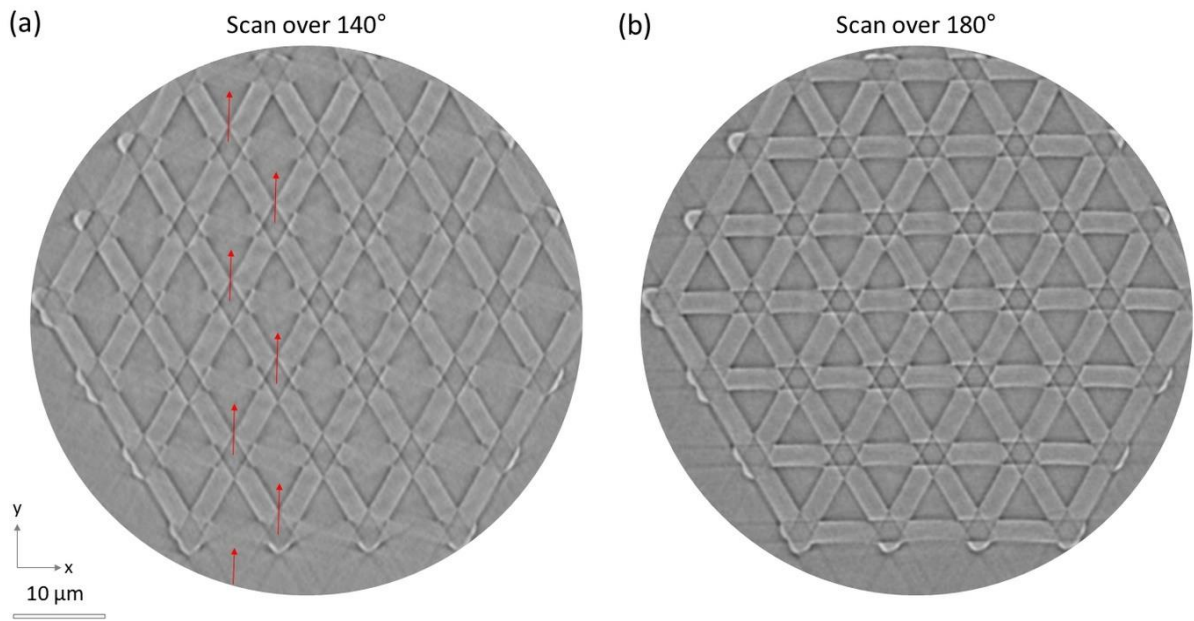

Figure 3: XY view of a) Sample A scanned from -70° to +70° (projections acquired over 140°). Red arrows indicate the horizontal missing beams rows. b) Sample A scanned from -90° to +90° (projection acquired over 180°). Both samples were scanned in Zernike phase contrast mode and reconstructed using the software Zeiss Scout and Scan™ Control System Reconstructor. The diamonds (a) and stars (b) in the nodes of the structure are reconstruction artifacts.

To improve the resolution at the reduced angular scanning range of 140° we increased the number of projections from 901 to 1601. This did not alter the reconstruction results. Rotating the sample to illuminate different angles with respect to the microstructure (Figure 4) improved the results. The XY views for the sample placed in 0° (typical position), 45° and 90° with respect to the anvil and scanned in phase and absorption contrast modes are shown in

Figure 3. We found that turning the sample to 45° gives a better reconstruction compared to 0°. With the sample positioned at 90°, the reconstruction of all beams is equivalent to the scan over 180° (Figure 3b).

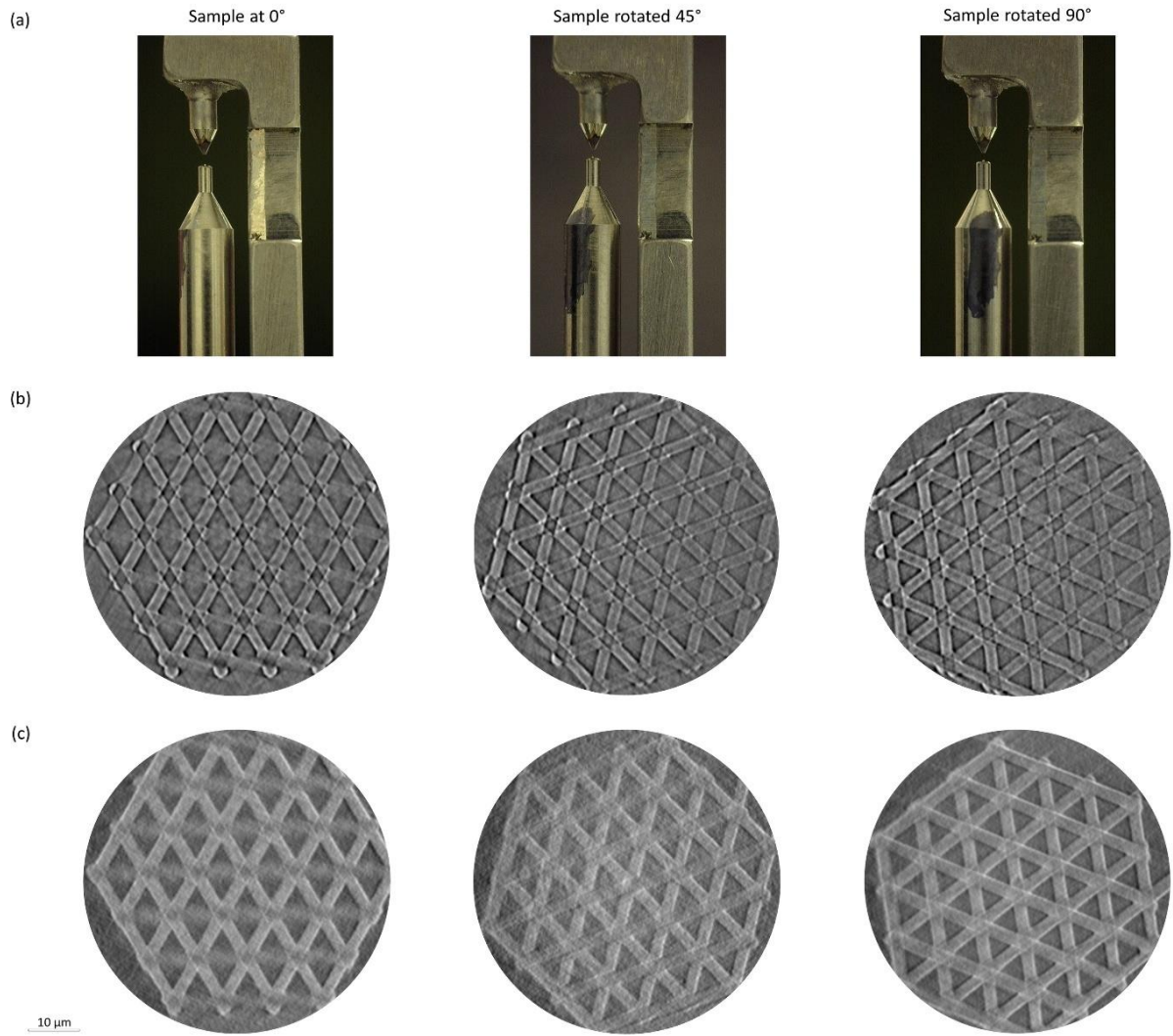

Figure 4: a) Position of the sample on the pin in relation to the anvil in the load stage (pen marking), corresponding to the typical position (0°) and the pin rotated 45° and 90°. XY views of sample A scanned from -70° to +70° with the sample placed into different angles in the stage, in (b) Zernike phase contrast and (c) absorption contrast.

For a better visualization of the difference in the reconstruction of the absorption contrast scans, a 5 µm line was drawn in a corresponding "horizontal beam" (Figure 5a) and the grayscale values were plotted as a function of the line distance (Figure 5b). The grayscale distribution in the reconstruction of the samples scanned over 180° and over 140° at position 90° are similar, with an expect increase in the intensity in the region corresponding to the "horizontal beam". For the sample scanned over 140° at position 0° and 45°, the grayscale shows a small variation over the measured distance. The grayscale intensity values for the samples scanned over 180° and over 140° at position 90° vary about 23.000 between the background and the "horizontal beams" regions, while for the reconstructions of the scans over 140° at positions 0° and 45°, are mostly uniform with grayscale values varying approximately 9.000 and 7.000, respectively.

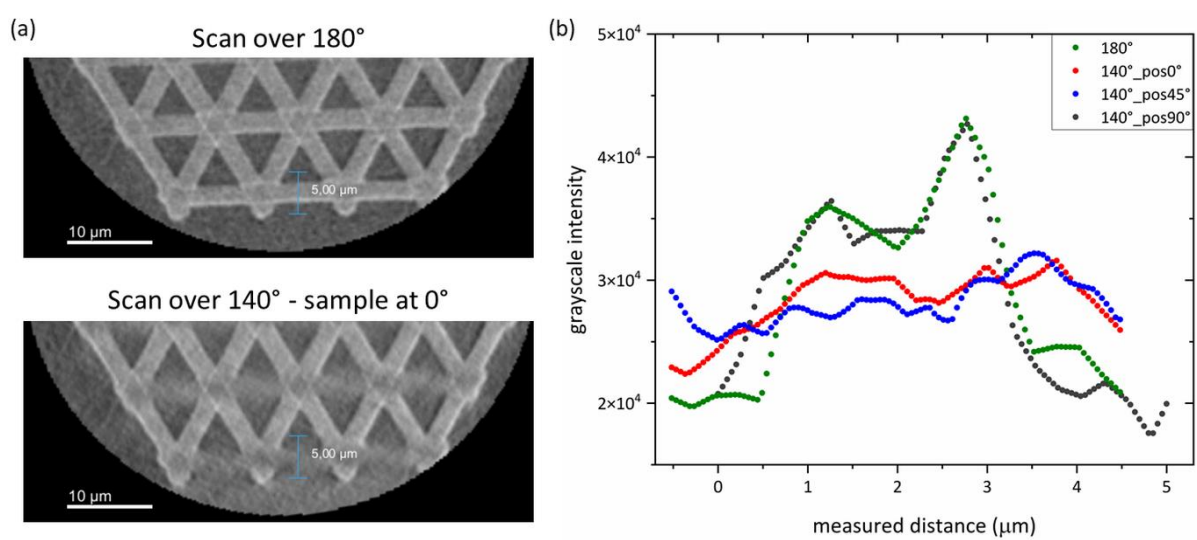

Figure 5: a) XY view of the sample scanned in absorption contrast over 180° and 140° with the sample positioned at 0°. A 5 µm line was drawn in a "horizontal beam" to extract the grayscale values of the region. b) Grayscale values distribution plotted as a function of the measuring distance from (a). Scans with an accurate reconstruction create a peak of grayscale distribution, while the scans with missing "horizontal beams" have a more uniform distribution of values.

With the aim of specifically characterizing the regions of the sample with lower information gain, sample B was scanned over 140° in the positions 0° and 90° with respect to the anvil. The results shown in Figure 6 demonstrate that the sample was only partially reconstructed. The region not reconstructed corresponds to an angular range of approximately 20°. In this specimen, the regions not directly illuminated by the X-ray beam due to the limited angle of projection acquisition are located on the left and right side of the sample. When rotating the sample by 90°, plates that were not reconstructed are now visible (symbols marked with a red circle in Figures 6 a and b).

In order to evaluate the influence of the density of the sample in such a setup, sample B was coated with 100 nm of $Al_2O_3$. Figure 7 shows that the coating led to an improved reconstruction of the sample. While in the polymeric sample two bars are missing in the reconstruction (Figure 7 a and c), in the coated sample only one bar is not fully reconstructed (Figure 7 b and d).

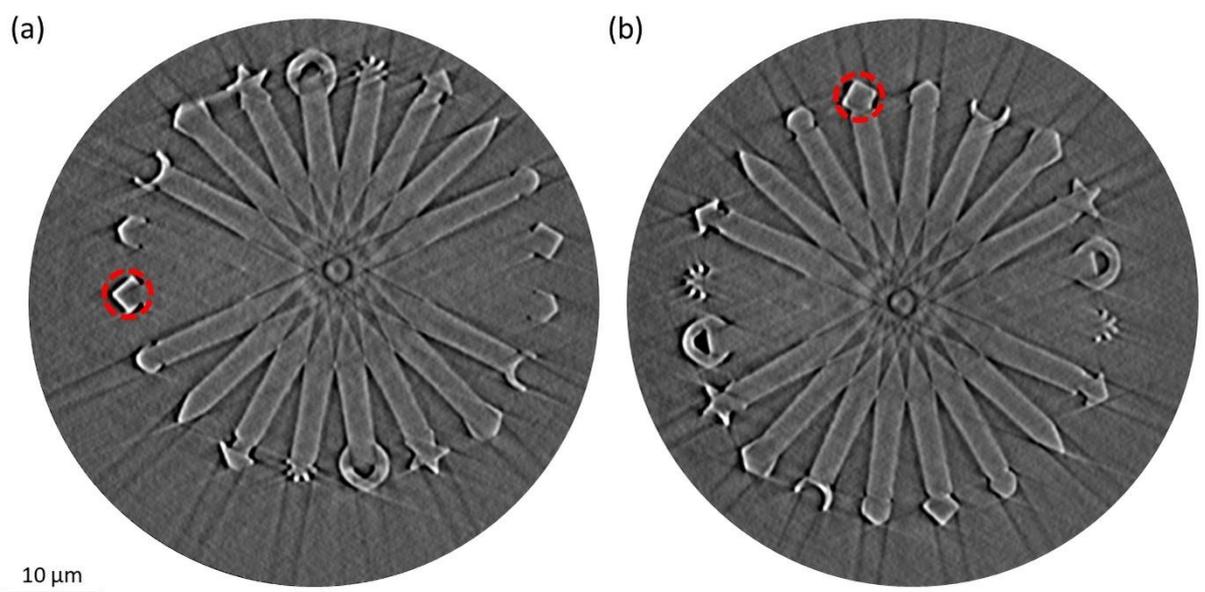

Figure 6: Sample B scanned over 140° positioned at 0° (a) and 90° (b) with respect to the anvil. The bars that are oriented in a certain angular position are not reconstructed, while the symbols are visible. The plate not visible in (a) (red circle) is visible when rotating the sample in 90°. These reconstructions are made from phase contrast imaging.

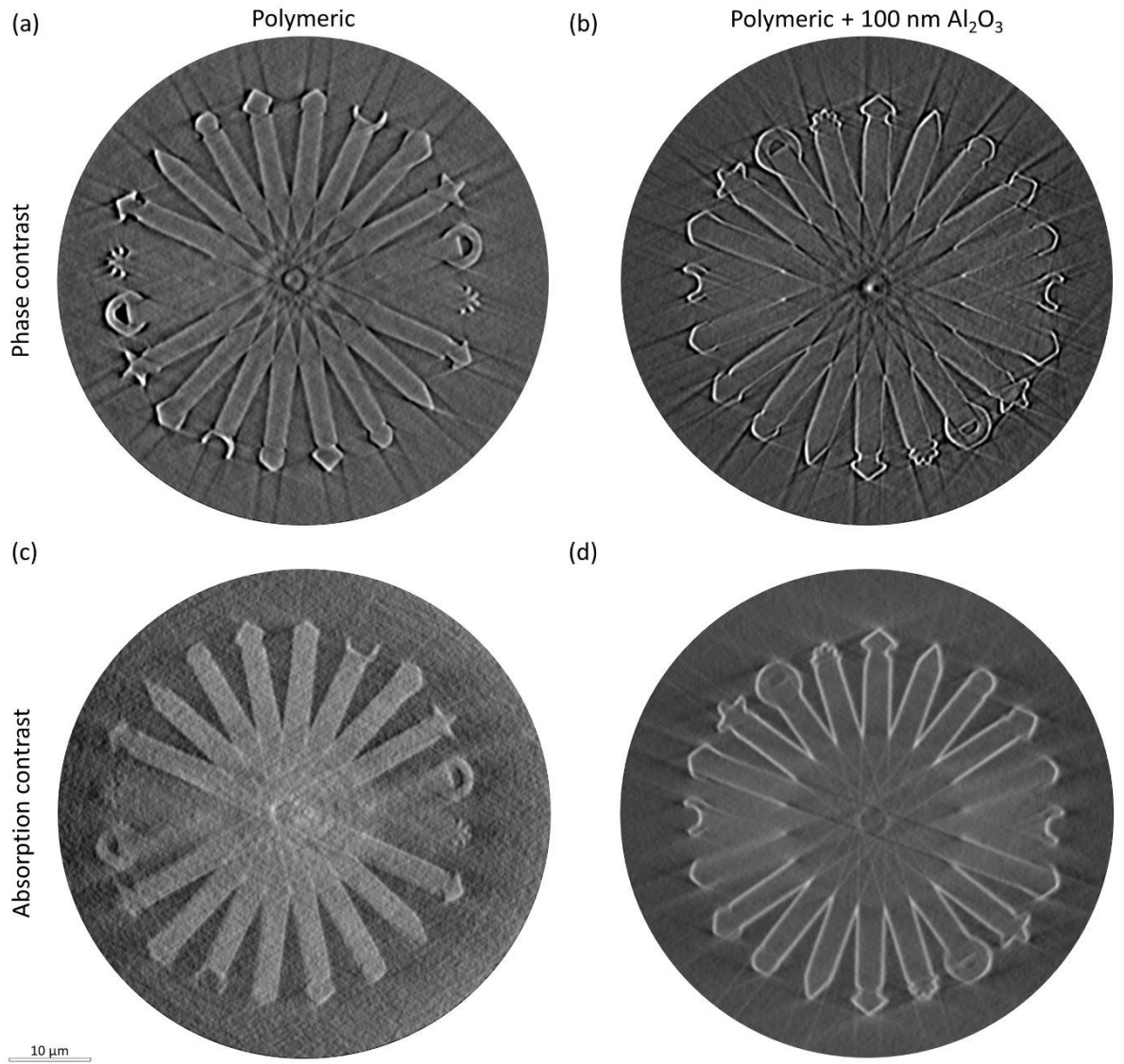

Figure 7: Sample B scanned over 140° in Zernike phase contrast (a and b) and absorption contrast (c and d) before (left – a and c) and after (right – b and d) coating the sample with 100 nm of $Al_2O_3$. The coating increased the contrast and improved the reconstruction, although did not eliminate the problem, as one of the plates is still not reconstructed.

## 4. Discussion

The low contrast of polymeric structures or metamaterials combined with the limited angle of scan imposed by the mechanical *in situ* load stage of the nanoCT causes the suppression of data during the reconstruction (Figure 3). The missing 40° sector of scanning range leads to an angular range of about 20° in which the surfaces within the structures or other weak contrast features are blurred or completely invisible. This holds for both absorption contrast and Zernike phase contrast (Figure 4) and cannot be improved by increasing the number of scans within the limited angle range of 140°. It is important to note that the missing angular wedge is not an effect of direct "shadowing" by the anvil but of the missing low angle illumination in certain directions combined with the low contrast of the polymer. This can be concluded from the remaining visibility of the symbols at the end of the missing bars in the poorly reconstructed region of the sample (Figure 6) and from the lack of visibility of the horizontal bars in the lattice structure (Figure 3). Due to the low contrast of the polymeric sample, the illumination at only large angles apparently is insufficient to reconstruct bars in the missing illumination directions. This suggests two possible routes along which the reconstruction problem might be tackled. First, one can attempt to enhance contrast by coating the sample and second one can try to avoid the alignment of structural features e.g. in lattice structures or metamaterials within this missing angular space.

Increasing the absorption contrast of the polymeric materials by coating them with 100 nm of $Al_2O_3$ was indeed sufficient to increase the contrast significantly. The absorption length for the X-ray energy of 5.4 keV for $Al_2O_3$ is 25 μm, while it is about 16 times larger [15] for the IP-Dip photoresist ($CH_2N_{0.001}O_{0.34}$, solid density 1.2 g/cm³) [16], corresponding to an absorption depth of 416.8 μm. The thin coating improves reconstruction results but does not completely eliminate the problem as one plate is still not reconstructed in Figure 6. Furthermore, coating a sample of course changes its mechanical properties [17] and therefore this is not really an option for most *in situ* mechanical testing experiments.

Metamaterials and other artificially structured materials are often built from beams or plates composed into lattice structures or geometrically simple building blocks [18]. In many cases, like in our example sample A, the elements are connected at angles larger than 45°. This leaves an angular space in the structure to which no surfaces and no structural elements are aligned. The horizontal missing beams in sample A correspond to the directions of the missing X-ray illumination during the limited angle scan. Rotating the sample by 90° rotates all relevant surfaces by about 30° to the original illumination direction. This creates a more even X-ray illumination for all the projections and leads to a complete reconstruction of the data, which is almost as good as for a complete 180° scan.

## 5. Conclusions

Dealing with *in situ* setups may impose limitations on tomography experiments. Using the mechanical load stage for *in situ* imaging in the lab-based X-ray microscope Xradia 810 Ultra implies scanning the samples over a limited angle of 140°. Although for many specimens this limitation does not interfere with the reconstruction of the data using a filtered back projection algorithm, it is a problem for low contrast, polymeric structures and particularly for microlattices at certain angular positions. Due to the geometry and low contrast of the polymeric microlattices, parts of the sample (the "horizontal beams") are not reconstructed. Increasing the contrast of the sample through a thin layer of $Al_2O_3$ improves the reconstruction significantly but of course modifies the sample properties.

To resolve this issue in lattice structures without modifying the sample is to rotate the sample to an angle that allows for a more even X-ray illumination of the internal surfaces and avoiding beam orientations parallel to the missing angular wedge of the illumination. Accurate reconstruction of our tetrahedral sample was possible despite the limited angular scanning range when the sample was scanned at such position.

For such low contrast polymeric structures it is therefore advisable to scan the sample over both the complete 180° and the limited 140° range of the *in situ* setup before actually starting the *in situ* experiment to make sure that all relevant structural elements are visible.

**Acknowledgement**

This work was partly carried out with the support of the Karlsruhe Nano Micro Facility (KNMF, www.knmf.kit.edu), a Helmholtz Research Infrastructure at Karlsruhe Institute of Technology (KIT, www.kit.edu). The Xradia 810 Ultra (nanoCT) core facility was supported (in part) by the 3DMM2O - Cluster of Excellence (EXC-2082/1390761711). R.D., C.M.K., E.D.L., M.B., R.S and P.G. acknowledge the support by the Cluster of Excellence 3DMM2O (EXC_2082/1-390761711) funded by the German Research Foundation (DFG). The work of E.D.L. has been supported by a postdoctoral research fellowship of the Alexander von Humboldt Foundation.